\documentstyle[preprint,aps]{revtex}

\begin{document}

\title{Momentum Imparted By Gravitational Waves with Spherical Wavefronts}

\author{M. Sharif \thanks{Permanent Address: Department of
Mathematics, Punjab University, Quaid-e-Azam Campus Lahore-54590,
PAKISTAN, e-mail: hasharif@yahoo.com}\\ Department of Physics,
Hanyang University, Seoul 133-791, KOREA}

\maketitle

\begin{abstract}
{\it A different approach has been used to evaluate the momentum
imparted by the gravitational waves with spherical wavefronts. It
is shown that the results obtained for momentum coincide with
those already available in the literature.}
\end{abstract}

\newpage

General Relativity (GR) predicts the existence of gravitational
waves as solutions of Einstein's field equations [1]. Since these
waves, by definition, have zero stress-energy tensor, a natural
question arises: How can these solutions of field equations
represent waves if they carry no energy? Essentially this problem
arises because energy is not well defined in GR. For energy to be
well defined the metric must have a time like isometry so as to
allow time translational invariance. This will not generally be
true. Infact, spacetimes for which it is true are static whereas
gravitational wave solutions must be non-static. Thus energy is
not well defined for spacetimes containing gravitational waves.
The problem was resolved by Ehlers and Kundt [2], Pirani [3] and
Weber and Wheeler [4] by considering a sphere of test particles in
the path of the waves. They showed that these particles acquired a
constant momentum from the waves. However, the procedure [5] to
evaluate momentum is quite complicated. An operational procedure
called the extended pseudo-Newtonian $(e\psi N)$ formalism [6]
embodying the same principle has been used to express the
consequences of relativity in terms of the Newtonian concept of
gravitational force. We have provided a closed form expression for
the momentum imparted to test particles in arbitrary spacetimes
[7,8]. By using this procedure for the plane and cylindrically
gravitational waves [7], we have already obtained physically
reasonable results coinciding with the earlier available results
in the literature. In this paper we apply the same procedure to
evaluate momentum imparted by gravitational waves with spherical
wavefronts. We find that the momentum expression for such waves
evaluated by this procedure coincides with that already evaluated
by using M$\ddot{o}$ller's prescription [9]. We shall not discuss
the $e\psi N$ formalism in any detail as it is available elsewhere
[6-8] rather we shall suffice to give its essential points.

The relativistic analogue of the Newtonian gravitational force
called the $\psi N$ gravitational force, is defined as the
quantity whose directional derivative along the accelerometer,
placed along the principal direction, gives the extremised tidal
force and which is zero in the Minkowski space. The $e\psi N$
force, $F_\mu$, is given as
\begin{equation}
F_0=m\left[\{\ln
(A/\sqrt{g_{00}})\}_{,0}-g^{ij}_{,0}g_{ij,0}/4A\right], \quad
F_i=m(\ln\sqrt{g_{00}})_{,i},\quad (i,j=1,2,3),
\end{equation}
where $A=(\ln \sqrt{-g})_{_{,0}},\quad g=det(g_{_{ij}})$. This
force formula depends on the choice of frame, which is not
uniquely fixed.

The quantity whose proper time derivative is $F_\mu$ called the
momentum four-vector for the test particle. Thus the momentum
four-vector, $p_\mu$, is

\begin{equation}
p_\mu=\int F_\mu dt,\quad (\mu=0,1,2,3).
\end{equation}
The spatial components of this vector give the momentum imparted to test
particles as defined in the preferred frame (in which $g_{_{0i}}=0)$.

The gravitational waves with spherical wavefronts are given by the
line element of the form [10]
\begin{equation}
ds^2=e^{-M}(dt^2-d\rho^2)-e^{-U}(e^Vdz^2+e^{-V}d\phi^2),
\end{equation}
where the metric functions $U,V$ and $M$ depend on the coordinates
$t$ and $\rho$ only. Einstein's vacuum field equations imply that
$e^{-U}$ satisfies the wave equation
\begin{equation}
(e^{-U})_{tt}-(e^{-U})_{\rho\rho}=0,
\end{equation}
and that $V$ satisfies the linear equation
\begin{equation}
V_{tt}-U_tV_t-V_{\rho\rho}+U_{\rho}V_{\rho}=0.
\end{equation}
The remaining equations for $M$ are
\begin{equation}
U_{tt}-U_{\rho\rho}=\frac 12
(U_t^2+U_{\rho}^2+V_t^2+V_{\rho}^2)-U_tM_t-U_{\rho} M_{\rho}=0,
\end{equation}
\begin{equation}
2U_{t\rho}=U_tU_{\rho}-U_tM_{\rho}-U_{\rho}M_t+V_tV_{\rho}.
\end{equation}
It is well known that, if Eqs.(4) and (5) are satisfied, the
Eqs.(6) and (7) are automatically integrable.

Using Eqs.(1) and (3), it follows that the $e\psi N$ force turns
out to be
\begin{equation}
F_0=m[\dot{U}+\frac{\ddot{M}+2\ddot{U}}{\dot{M}+2\dot{U}}-
\frac{3\dot{U}^2+\dot{V}^2}{\dot{M}+2\dot{U}}],\quad
F_1=-m\frac{M'}{2},\quad F_2=0=F_3.
\end{equation}
The corresponding four-vector momentum will become
\begin{equation}
p_0=m[U+\ln(\dot{M}+2\dot{U})-\int
\frac{3\dot{U}^2+\dot{V}^2}{\dot{M}+2\dot{U}}dt]+f_1(\rho),
\end{equation}
\begin{equation}
p_1=-\frac{m}{2}\int M'dt+f_2(\rho),\quad p_2=constant=p_3.
\end{equation}
where dot denotes differentiation with respect to time and prime
with respect to $\rho$, $f_1$ and $f_2$ are arbitrary functions of
$\rho$. Eqs.(9) and (10) provide the general expression of the
momentum four-vector for the gravitational waves with spherical
wavefronts. As we are interested in evaluating the momentum
imparted  by gravitational waves we need to calculate the term
$p_1$. For this purpose, we require the value of $M$.

The background region ($t<\rho$, Minkowski) is described by the
solution \\ $U=-\ln t-\ln\rho,\quad V=\ln t-\ln\rho$ and $M=0$.
Substituting these values in Eqs.(9) and (10), we have
\begin{equation}
p_0=m\ln(-2/\rho)+f_1(\rho),\quad p_i=constant.
\end{equation}
The quantity $p_0$ can be made zero by choosing
$f_1(\rho)=-m\ln(-2/\rho)$ and the momentum term $p_i$ will be
zero for a particular choice of an arbitrary constant as zero.
Thus the four-vector momentum vanishes in the background region
(Minkowski) as was expected.

The solution on the wavefront ($t=\rho$) can be written in the
form\\ $U=-2\ln t,\quad V=0,\quad M=0$. Using these values in
Eqs.(9) and (10), it follows that
\begin{equation}
p_0=m\ln(-4)+f_1(\rho),\quad p_i=constant.
\end{equation}
We see that the momentum turns out to be constant which can be
made zero if we choose constant as zero.

The solution in the wave region ($t>\rho$) can be found by solving
Eqs.( 4) and (5) and is given in the form [10,11]
\begin{equation}
U=-\ln t-\ln\rho,\quad V=\ln t-\ln\rho+\tilde{V}(t,\rho).
\end{equation}
With this, the remaining equations for M become, for $t>\rho$
\begin{eqnarray}
M_t+M_\rho=(\frac{t-\rho}{t+\rho})(\tilde{V}_t+\tilde{V}_\rho)
-\frac{t\rho}{2(t+\rho)}(\tilde{V}_t+\tilde{V}_\rho)^2,\\\nonumber
M_t-M_\rho=(\frac{t+\rho}{t-\rho})(\tilde{V}_t-\tilde{V}_\rho)
+\frac{t\rho}{2(t-\rho)}(\tilde{V}_t-\tilde{V}_\rho)^2.
\end{eqnarray}
We now make observation that the characteristic boundary condition
(13) can be satisfied by the infinite set of particular solutions
of self-similar type
\begin{equation}
\tilde{V}(t,\rho)=(t\rho)^kH_k(\frac{t^2+\rho^2}{2t\rho}),
\end{equation}
where $k$ is an arbitrary real or complex number, and the
functions $H_k(\frac{t^2+\rho^2}{2t\rho})$ satisfy some ordinary
linear equation which can be reduced to a hypergeometric equation
and the condition $H_k(1)=0$. This form of solution has been used
in [12,13] in a different context. As an explicit example we may
consider the case of a single component
\begin{equation}
\tilde{V}(t,\rho)=a_k(t\rho)^kH_k(\frac{t^2+\rho^2}{2t\rho})
\end{equation}
for some $k\geq\frac 12$ and constant $a_k$. In this case, $M$ is
given by
\begin{equation}
M=\frac{1}{2k}a_k(t^2-\rho^2)(t\rho)^{k-1}H_{k-1}-\frac{1}{2k}
(t\rho)^{2k}a^2_k[k^2H^2_k-\frac{(t^2-\rho^2)^2}{4t^2\rho^2}H^2_{k-1}].
\end{equation}
Notice that the dimension of $a_k$ is $L^{-2k}$. If we use this
value of $M$ after taking its derivative in Eq.(10), it may not be
possible to perform its integration. For the purpose of
simplicity, we take a special case when $k=1$ for which $M$ takes
the form
\begin{equation}
M=\frac{1}{2}a_1(t^2-\rho^2)H_0-\frac{1}{2}(t\rho)^2a^2_1[H^2_1
-\frac{(t^2-\rho^2)^2}{4t^2\rho^2}H^2_0],
\end{equation}
where $H_0=\ln(t/\rho),\quad H_1=\frac
12[(t/\rho+\rho/t)\ln(t/\rho)-(t/\rho-\rho/t)]$. After taking
derivative of $M$ with respect to $\rho$, we substitute it in
Eq.(10) and after a tedious integration, we obtain
\begin{eqnarray}
p_1=m\frac{a_1}{4}[\frac 23(t^3/\rho)-6t\rho+4t\rho\ln(t/\rho)
+a_1\{\frac{1}{5}(t^5/\rho)+\frac{2}{27}t^3\rho+5t\rho^3\\\nonumber
-4(\frac{5}{9}t^3\rho+t\rho^3)\ln(t/\rho)+\frac{4}{3}t^3\rho(\ln(t/\rho))^2\}]
+f_2(\rho).
\end{eqnarray}
This gives the momentum imparted to test particles by
gravitational waves with spherical wavefronts. The quantity $p_1$
can be made zero for the $t\rightarrow 0$ limit by choosing
$f_2=0$. However, it immediately indicates the presence of
singularity when $\rho=0$ and this singularity at $\rho=0$ acts as
a source of the gravitational waves inside the wave region. This
coincides with the result evaluated by using M$\ddot{o}$ller's
prescription [9]. This is a physically reasonable expression for
the momentum imparted by gravitational waves. The interpretation
of $p_0$ in the $e\psi N$ formalism is given elsewhere [14]. It
can also be shown that, near the wavefront as $t\rightarrow\rho$
\begin{equation}
\tilde{V}\sim a_k(t+\rho)^{-1}(t-\rho)^{1+2k},\quad M\sim
(\frac{1+2k}{2k})a_k(t-\rho)^{2k}.
\end{equation}
Using Eqs.(10) and (20), it follows that
\begin{equation}
p_1=\frac{1}{4k}m(1+2k)a_k(t-\rho)^{2k}+f_2(\rho).
\end{equation}
This gives the momentum near the spherical wavefront. We see that
for a particular choice of $f_2$, it reduces to the momentum
expression given by Eq.(12) on the wavefront. We remark that our
results exactly coincide with those evaluated by using
M$\ddot{o}$ller's prescription for the background region and on
the wavefront. For the wave region, these two can be equated for a
particular choice of an arbitrary function $f_2$. We have seen
that in all the three cases we obtain a physically reasonable
expression for the momentum.

\newpage

\begin{description}
\item  {\bf Acknowledgment}
\end{description}

The author would like to thank Prof. Chul H. Lee for his
hospitality at the Department of Physics and Korea Scientific and
Engineering Foundation (KOSEF) for postdoc fellowship at Hanyang
University Seoul, KOREA.

\vspace{2cm}

{\bf \large References}

\begin{description}

\item{[1]} Misner, C.W., Thorne, K.S. and Wheeler, J.A.: {\it Gravitation} (W.H. Freeman,
San Francisco, 1973).

\item{[2]} Ehlers, J. and Kundt, W.: {\it Gravitation: An Introduction to Current
Research}, ed. L. Witten (Wiley, New York, 1962)49.

\item{[3]} Pirani, F.A.E.: {\it Gravitation: An Introduction to Current
Research}, ed. L. Witten (Wiley, New York, 1962)199.

\item{[4]} Weber, J. and Wheeler, J.A.: Rev. Mod. Phy. {\bf 29}(1957)509.

\item{[5]} Weber, J.: {\it General Relativity and Gravitational Waves}, (Interscience,
 New York, 1961).

\item{[6]} Qadir, A. and Sharif, M.: Nuovo Cimento {\bf
107B}(1992)1071;\\ Sharif, M.: Ph.D. Thesis, Quaid-i-Azam
University (1991).

\item{[7]} Qadir, A. and Sharif, M.: Physics Letters {\bf A
167}(1992)331.

\item{[8]} Sharif, M.: Astrophys. Space Sci. {\bf 253}(1997)195;
{\bf 262}(1999)297.

\item{[9]} M$\ddot{o}$ller, C.: Ann. Phys. (NY) {\bf 4}(1958)347; {\bf 12}(1961)118.

\item{[10]} Alekseev, G.A. and Griffths, J.B.: Class. Quantum Grav. {\bf
12}(1995)L13.

\item{[11]} Alekseev, G.A. and Griffths, J.B.: Class. Quantum Grav. {\bf
13}(1996)2191.

\item{[12]} Griffths, J.B.: Class. Quantum Grav. {\bf 10}(1993)975.

\item{[13]} Alekseev, G.A. and Griffths, J.B.: Phys. Rev. {\bf D
52}(1995)4497.

\item{[14]} Qadir, Asghar, Sharif, M. and Shoaib, M.: Nuovo Cimento {\bf 115B}(2000)419.

\end{description}

\end{document}